\documentclass[aps,prl,reprint,a4paper,twocolumn,showpacs,floatfix]{revtex4}
\usepackage{graphicx}
\usepackage{amsfonts}
\usepackage{amsmath}
\usepackage{amssymb}
\usepackage{bm}
\usepackage{dcolumn}
\usepackage{hyperref}
\usepackage[usenames]{color}

\hypersetup{pdfborder=0 0 0,colorlinks=true,
citecolor=blue,linkcolor= blue}
\begin{document}
\title{Polarity-induced oxygen vacancies at LaAlO$_3|$SrTiO$_3$ interfaces}
\author{Zhicheng Zhong}
\affiliation{Faculty of Science and Technology and MESA$^+$
Institute for Nanotechnology, University of Twente, P.O. Box 217,
7500 AE Enschede, The Netherlands}
\author{P. X. Xu}
\affiliation{Faculty of Science and Technology and MESA$^+$
Institute for Nanotechnology, University of Twente, P.O. Box 217,
7500 AE Enschede, The Netherlands}
\author{Paul J. Kelly}
\affiliation{Faculty of Science and Technology and MESA$^+$
Institute for Nanotechnology, University of Twente, P.O. Box 217,
7500 AE Enschede, The Netherlands}

\date{\today}

\begin{abstract}
Using first-principles density functional theory calculations, we find a strong position and thickness dependence of the formation energy of oxygen vacancies in LaAlO$_3|$SrTiO$_3$ (LAO$|$STO) multilayers and interpret this with an analytical capacitor model. Oxygen vacancies are preferentially formed at $p$-type SrO$|$AlO$_2$ rather than at $n$-type LaO$|$TiO$_2$ interfaces; the excess electrons introduced by the oxygen vacancies reduce their energy by moving to the $n$-type interface. This asymmetric behavior makes an important contribution to the conducting (insulating) nature of $n$-type ($p$-type) interfaces while providing a natural explanation for the failure to detect evidence for the polar catastrophe in the form of core level shifts.
\end{abstract}

\pacs{68.35.-p,  68.35.Ct,   73.20.-r}
%
%
%
%
%
%
%
%
%
%
%
%
%

\maketitle

{\em \color{red} Introduction:} Extremely high carrier mobilities
have recently been observed when interfaces consisting of LaO and
TiO$_2$ layers are formed between insulating LaAlO$_3$ and
SrTiO$_3$ perovskites \cite{Ohtomo:nat04}. Even though the
physical origin of this metallic behavior is still under debate
\cite{Nakagawa:natm06,Siemons:prl07,Herranz:prl07,Kalabukhov:prb07,Yoshimatsu:prl08,Kalabukhov:prl09},
most experimental
\cite{Huijben:natm06,Thiel:sc06,Brinkman:natm07,Reyren:sc07,Vonk:prb07,Willmott:prl07,Basletic:natm08,Cen:natm08,Caviglia:nat08,Salluzzo:prl09,Sing:prl09,Huijben:am09}
and theoretical
\cite{Park:prb06,Pentcheva:prl09,Schwingenschlogl:epl09,Chen:prb09,Popovic:prl08,Janicka:prl09,Bristowe:prb09,Zhong:epl08,Kumar:prb08}
studies have reached a consensus that the so-called {\em polarity
discontinuity} between these materials plays a crucial role; in
the absence of any relaxation mechanism, alternate stacking of
positively (LaO$^+$) and negatively (AlO$_2^-$) charged layers on
the non-polar STO substrate would give rise to a huge effective
internal electric field, leading to a divergence of the
electrostatic potential with increasing thickness of LAO. Three
mechanisms have been suggested to avoid this instability: charge
transfer \cite{Nakagawa:natm06}, atomic relaxation
\cite{Vonk:prb07,Huijben:am09,Park:prb06,Pentcheva:prl09,Schwingenschlogl:epl09,Chen:prb09},
or the creation of oxygen vacancies
\cite{Herranz:prl07,Kalabukhov:prb07} and other defects
\cite{Kalabukhov:prl09,Willmott:prl07}.

The first mechanism refers \cite{Nakagawa:natm06} to the transfer of electrons from a surface AlO$_2$ layer to the interface TiO$_2$ layer by the internal electric field. The excess charge at the interface balances the polar discontinuity and leads to conducting behavior of the interface. This mechanism is strongly supported by the observation of an insulator-metal transition induced by either an external electric field or by increasing the thickness of the LaAlO$_3$ layer \cite{Thiel:sc06}. However, direct experimental evidence of charge transfer, in the form of core level shifts, has
not yet been found \cite{Yoshimatsu:prl08}. The insulating behavior of the $p$-type interface \cite{Ohtomo:nat04} is also not readily accommodated in this picture.

The second mechanism, atomic relaxation in the presence of the
internal electric field, that is analogous to the buckling of
Ti-O-Ti chains in an external field in SrTiO$_3$, has been
discussed by a number of authors
\cite{Vonk:prb07,Huijben:am09,Park:prb06,Pentcheva:prl09,Schwingenschlogl:epl09,Chen:prb09}.
It can eliminate the diverging potential by introducing a
compensating electric field. A third way to resolve the polar
instability is to introduce defects at interfaces.  Oxygen
vacancies (and other defects) created during the growth of LAO on
STO are invoked by Herranz {\em et al.} \cite{Herranz:prl07} and
Kalabukhov {\em et al.} \cite{Kalabukhov:prb07} to understand the
high mobility carriers. The long-relaxation time of the
electric-field-induced insulator-metal transition
\cite{Cen:natm08} suggests the possibility of interface defect
diffusion.

While the first two mechanisms have received much theoretical
attention
\cite{Park:prb06,Zhong:epl08,Popovic:prl08,Schwingenschlogl:epl09,Chen:prb09,Pentcheva:prl09,Janicka:prl09,Bristowe:prb09},
the relationship between the creation of defects and polarity has
not been clarified theoretically. To demonstrate the coupling of
polarity and oxygen vacancy formation and throw some light on the
interplay with atomic relaxation and charge transfer, we calculate
from first-principles the formation energy of oxygen vacancies in
LAO$|$STO multilayers as a function of their location in the
multilayer.

\begin{figure}[t!]
\includegraphics[scale=0.5]{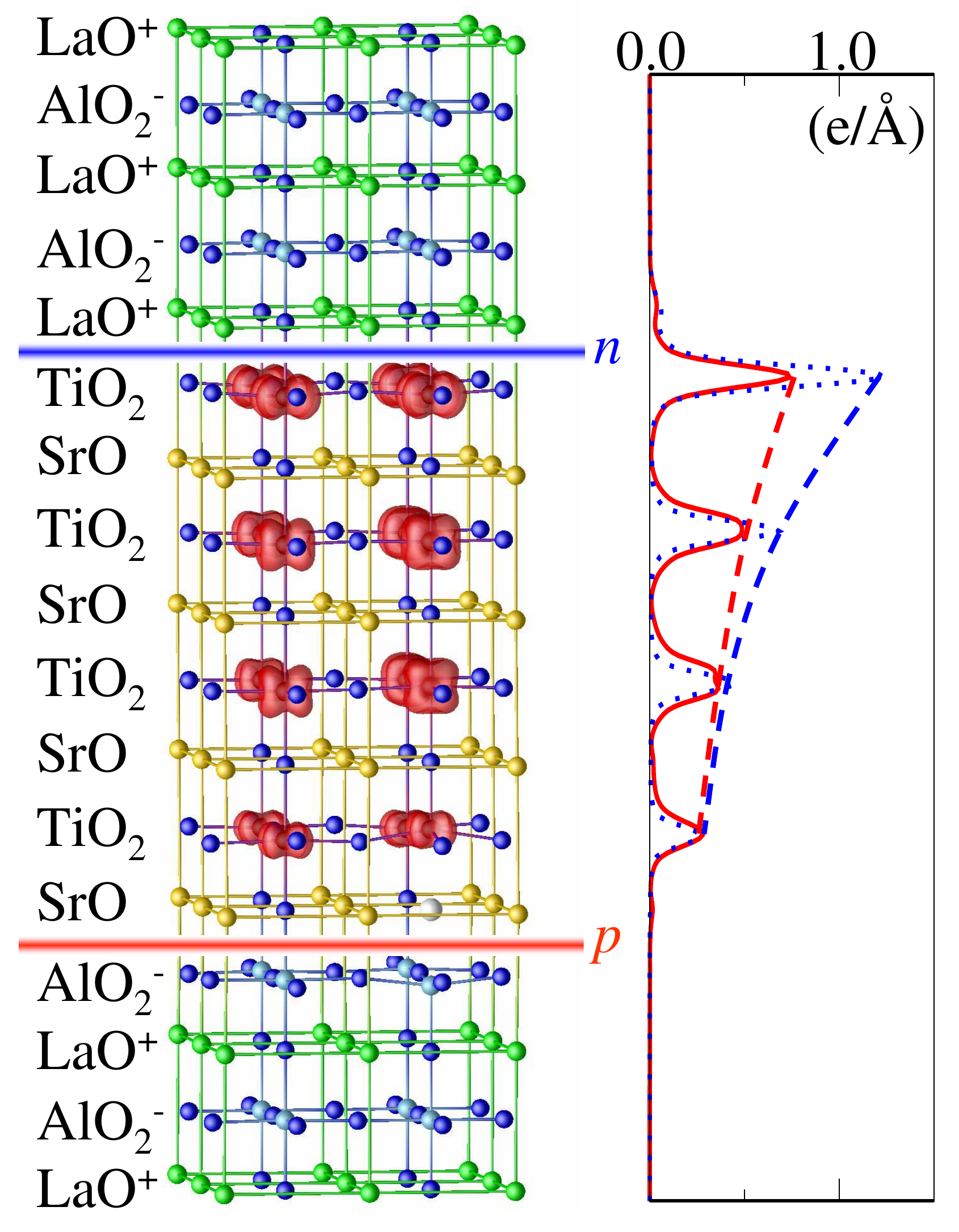}
\caption{Left panel: the unit cell of a $2\times2\times(4+4)$ LAO$|$STO multilayer with an oxygen vacancy at the $p$-type interface. Blue spheres represent oxygen atoms and the oxygen vacancy is marked by a white sphere. Charge density isosurfaces corresponding to a value of 0.015 $e / \AA^3$ for occupied states in the conduction bands are coloured red. Right panel: plane-averaged charge density as a function of $z$ for oxygen vacancies at $p$- (red) or $n$- (blue) type interfaces. }
\label{structure}
\end{figure}

{\em \color{red} Method:} We focus on ($m, m$) LAO$|$STO multilayers containing $m$ layers each of LAO and STO with alternating $p$- and $n$-type interfaces. Because samples are grown on STO substrates, we fix the in-plane lattice constant at the calculated equilibrium value of STO and calculate the out-of-plane lattice constant by minimizing the total energy of strained bulk LAO. Oxygen vacancies are modelled in a $2\times2$ lateral supercell and for each vacancy position all atoms are allowed to relax with the volume of the structure fixed. Most of the results reported below were obtained with the 159 atom
supercell depicted in Fig.~\ref{structure} containing a $p$ and $n$ interface. The periodically repeated single oxygen vacancy in a layer consisting of $2\times2$ unit cells should be compared to the $\sim$25\% oxygen vacancy concentration in a layer suggested by experiment \cite{Nakagawa:natm06}.
The local density approximation (LDA) calculations were carried out with the projector augmented wave method \cite{Blochl:prb94b} as implemented in the Vienna Ab-initio Simulation Package (VASP) \cite{Kresse:prb99}. A kinetic energy cutoff of 500~eV was used and the Brillouin zone of the 159 atom supercell was sampled with an $8\times8\times2$ k-point grid in combination with the tetrahedron method.

The concentration of vacancies in thermodynamic equilibrium is determined by the vacancy formation energy, $E_{\rm Ox}^{\rm Vac}$. Since we want to focus on how it depends on the position ($z$) of the defect for a charge neutral system, we express $E_{\rm Ox}^{\rm Vac}$ for an oxygen vacancy in an LAO$|$STO multilayer as the difference in total energies of supercells with and without a vacancy, $E_{\rm Ox}^{\rm Vac} = E_{\rm SC}^{\rm Vac} - E_{\rm SC}$ \cite{Zhang:prl91}. Because it is useful to relate this to the formation energy of an oxygen vacancy in bulk STO, $E_{\rm STO}^{\rm Vac}$, all of the vacancy formation energies discussed below are $\Delta E_{\rm Ox}^{\rm Vac} = E_{\rm Ox}^{\rm Vac} - E_{\rm STO}^{\rm Vac}$. Though $E_{\rm STO}^{\rm Vac}$ converges only slowly with the size of supercell \cite{Buban:prb04}, this does not alter the conclusions we will draw; $E_{\rm SC}^{\rm Vac}$ spans a range of $\pm 1.0$~eV so the effect of polarity is much larger than the uncertainty in $E_{\rm STO}^{\rm Vac}$ which is converged to $\pm 0.09$~eV.

\begin{figure}[t!]
\includegraphics[scale=0.50]{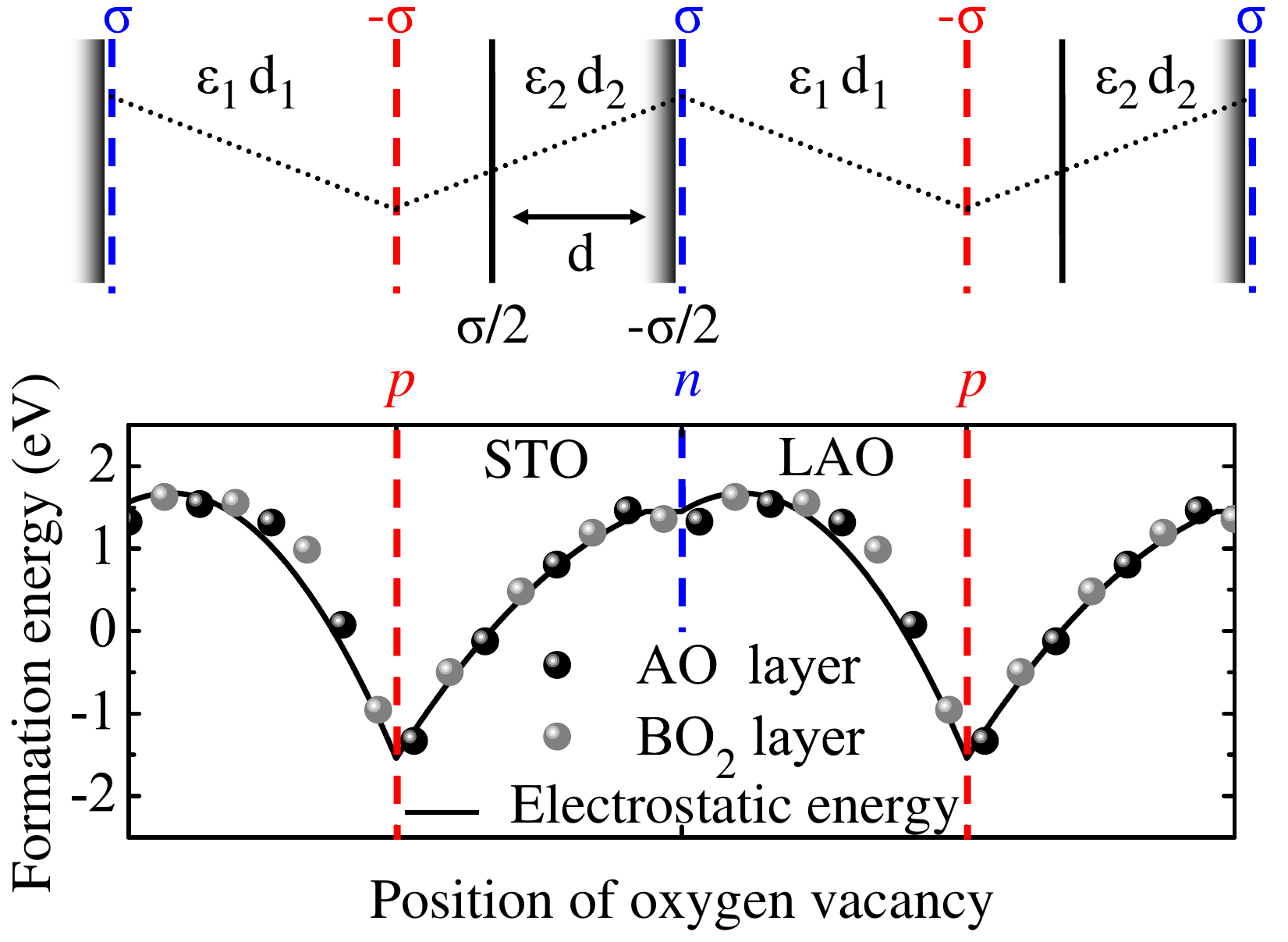}
\caption{Position dependence of the formation energy of an oxygen vacancy in a $2\times2\times(4+4)$ LAO$|$STO multilayer calculated from first principles without relaxation (symbols) and using an analytical capacitor model (solid line). $n$ and $p$ interfaces are indicated by vertical red and blue dashed lines. A schematic diagram of the capacitor model is shown in the upper panel. The electrostatic potential profile for the vacancy-free structure is shown as a dotted line. The vertical black line at a distance $d$ from the $n$ interface represents the oxygen vacancy layer. Two excess electrons are transferred to the TiO$_2$ layer at the $n$ interface (shaded grey line).  }
\label{layer}
\end{figure}

{\em \color{red} Unrelaxed results:} We begin by calculating $\Delta E_{\rm Ox}^{\rm Vac}(z)$ without atomic relaxation. The most striking feature of the results shown in the lower panel of Fig.~\ref{layer} is the asymmetry for forming a vacancy at the $n$ and $p$ interfaces. $\Delta E_{\rm Ox}^{\rm Vac}$ is lowest when the oxygen vacancy is at the $p$ interface, highest close to the $n$ interface and is nonlinear in $z$. It spans a range of about 3~eV between the two interfaces and differs in the LAO and STO layers. We can capture the essential behaviour of $\Delta E_{\rm Ox}^{\rm Vac}(z)$ in terms of a modified parallel-plate capacitor model.

{\em \color{red} Model:} The average electrostatic potential of defect-free LAO$|$STO multilayers, as probed by the energy levels of core states in LDA calculations, exhibits a simple symmetric triangular form as if the $n$ and $p$ interfaces were positively and negatively charged with charge density $\pm \sigma =\mp e/a^{2}$ where $a$ is the lattice constant of bulk STO. As sketched in the top panel of Fig.~\ref{layer}, the plates of the capacitor are separated by a thickness $d_1$ ($d_2$) of insulating LAO (STO) with dielectric constants $\varepsilon_1$ ($\varepsilon_2$) determined by the electronic polarization only in the absence of ionic relaxation. Such a model was recently used to describe the evolution of the dielectric properties of LAO$|$STO multilayers with increasing layer thickness resulting in an insulator-metal transition \cite{Bristowe:prb09}. Based on this simple capacitor model, the estimated internal electric fields are huge, $\frac{1}{\varepsilon }1.2\times 10^{11}V/m$ or $\sim 0.9$~V/unit cell in LAO, and the electrostatic potential (dotted line in Fig.~\ref{layer}) diverges with increasing thickness of LAO.

We extend this model to encompass the layer of vacancies constructed in our supercell approach. Because oxygen is divalent, removal of a neutral oxygen atom in a bulk insulating material such as LAO or STO leaves two excess electrons in the conduction band weakly bound to the oxygen vacancy. In an LAO$|$STO multilayer, the potential energy of the electrons in the internal electric field far exceeds this binding energy and the total energy can be reduced by moving the two electrons to the conduction band minimum at the $n$ interface leaving a sheet of positive charge at the oxygen vacancy plane a distance $d$ away. Assuming that the excess electrons are on Ti ions at the TiO$_2|$LaO interface, independent of where the oxygen vacancies were formed, we can calculate the $d$ dependent change in the electrostatic energy to be
$\frac{\sigma^2/\varepsilon _0}{d_1 \varepsilon_2 + d_2
\varepsilon_1} \{ -d_1 +
\frac{\varepsilon_1}{\varepsilon_2}(d_2-d) \}d$.
It comprises two parts: the energy to insert a sheet of positive charge density $\sigma/2$ in the LAO$|$STO capacitor background,
 $-\frac{2\sigma^2/\varepsilon_0}{d_1 \varepsilon_2 + d_2 \varepsilon_1} d_1d$,
and the potential energy of the positively and negatively charged sheets
$\frac{\sigma^2/\varepsilon_0}{d_1 \varepsilon_2 + d_2
\varepsilon_1} \{ d_1 + \frac{\varepsilon_1}{\varepsilon_2}
(d_2-d) \}d$.
The calculated LDA core level shifts of $\sim 0.9$V/unit cell can be used to estimate $\varepsilon_1 + \varepsilon_2 \sim 52$ leaving one free parameter in the model, the ratio $\varepsilon_2/\varepsilon_1$. Taking this to be 1.5 results in the solid curve in Fig.~\ref{layer}. The good fit of this simple model makes it clear that the internal fields induced by the polar layered structure can lower the formation energy of oxygen vacancies at the $p$ interface very substantially and that the origin of the asymmetry in formation energies in LAO and STO is the difference in their dielectric constants. The residual interaction between the field-ionized oxygen vacancies and electrons accounts for the nonlinear behavior of the formation energy.

\begin{figure}[t!]
\includegraphics[scale=0.55]{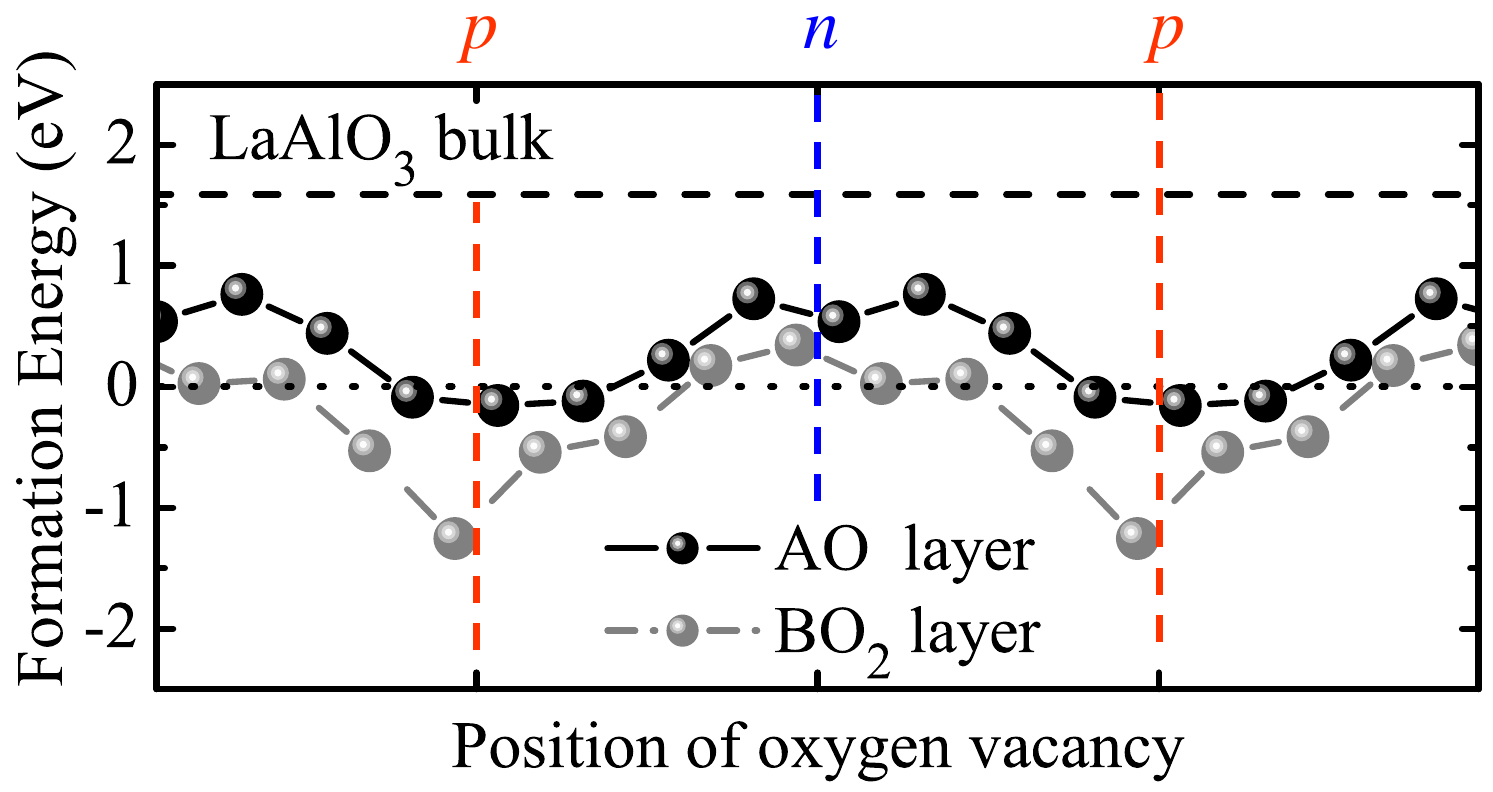}
\caption{Position dependence of the formation energy of an oxygen vacancy in a $2\times2\times(4+4)$ LAO$|$STO multilayer relative to that of a vacancy in bulk STO (horizontal dotted line) calculated from first principles with relaxation. Black and grey symbols are for vacancies in AO and BO$_2$ layers respectively. The formation energy of an oxygen vacancy in bulk LAO is shown as a dashed horizontal line.} \label{relaxed}
\end{figure}

Atomic relaxation can be expected to strongly suppress the polarity. Nevertheless, when our structures are fully relaxed, some essential features of $\Delta E_{\rm Ox}^{\rm Vac}(z)$ are unchanged, see Fig.~\ref{relaxed}. In particular, the formation energy has a minimum at the $p$ interface and a maximum close to the $n$ interface while the minimum formation energy is more than 1~eV lower than in bulk STO. Including atomic relaxation differentiates between vacancy formation in AO and BO$_2$ layers; the latter are energetically more favourable though the behaviour as a function of $z$ is essentially the same. The lower formation energy in BO$_2$ layers can be understood in terms of the types of relaxations possible within the constraints imposed by stacking the different layers in a multilayer. To simplify the discussion, we focus on the {\em less} favourable case of oxygen vacancies in AO layers so that our conclusions will also be applicable for BO$_2$-layer vacancies.

\begin{figure}[t!]
\includegraphics[scale=0.55]{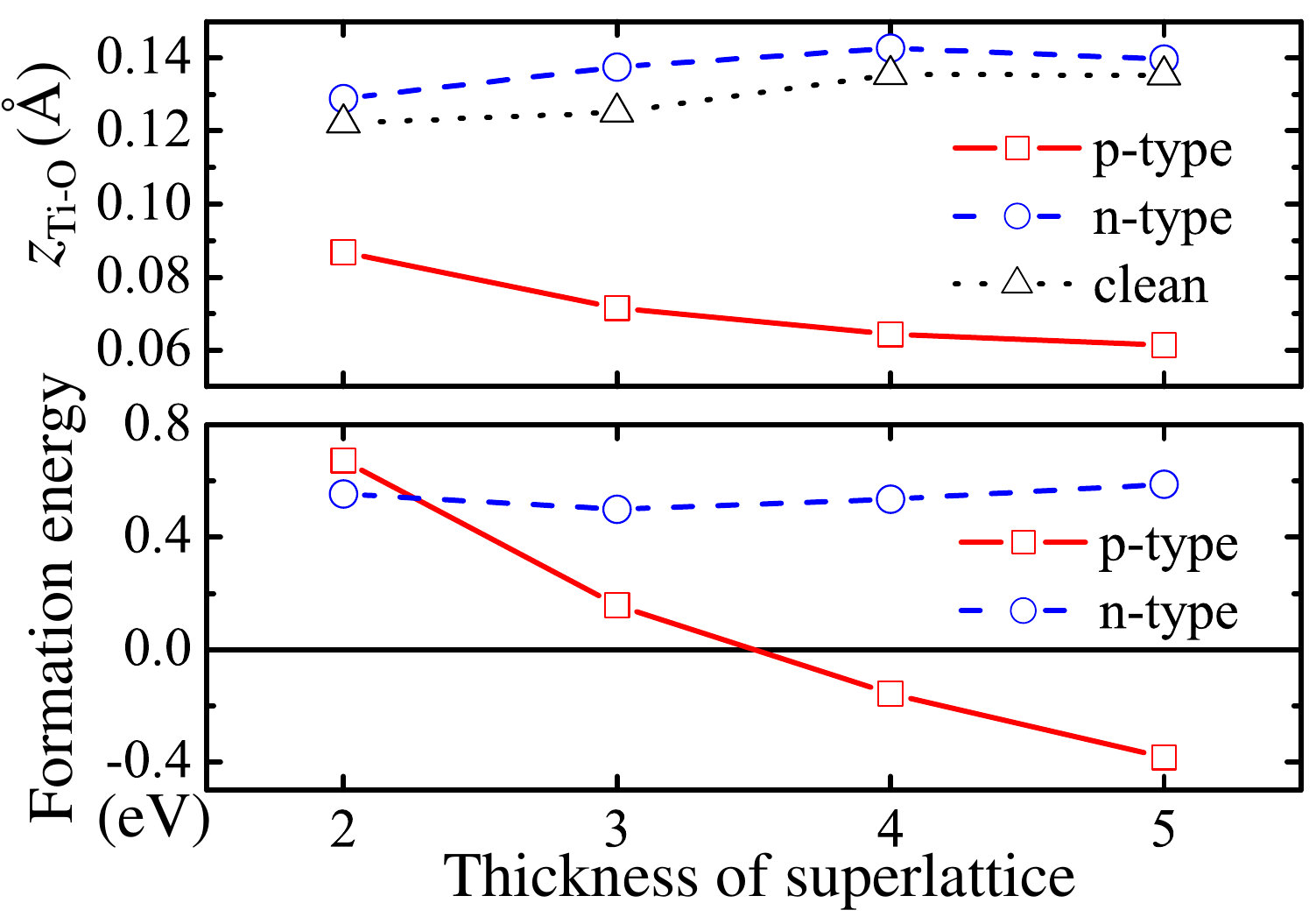}
\caption{Upper panel: projection of the Ti-O-Ti separation along the $z$ direction due to buckling for clean interfaces ($\triangle$) and when vacancies are formed at the $p$ ($\Box$) and at the $n$ ($\bigcirc$) interface. Lower panel: formation energy of an oxygen vacancy as a function of the multilayer thickness for $p$ and $n$ interfaces in a $2\times2\times(m+m)$ LAO$|$STO multilayer with relaxed structure.}
\label{criticalthickness}
\end{figure}

{\em \color{red} Critical thickness:} For a capacitor with fixed charge density $\sigma$, the electric field is constant and the electrostatic potential increases as the plate separation (LAO thickness) is increased, a feature that is supported by both experimental \cite{Thiel:sc06,Cen:natm08} and theoretical \cite{Park:prb06,Pentcheva:prl09,Schwingenschlogl:epl09,Chen:prb09,Bristowe:prb09} studies. Because of its dependence on the electrostatic potential, we expect the formation energy of oxygen vacancies to depend on the multilayer thickness. Since the minimum and maximum formation energies occur at or close to the $p$ and $n$ interfaces, respectively, we focus on these formation energies. We further assume equal thicknesses $m$ of the STO and LAO layers and plot $\Delta E_{\rm Ox}^{\rm Vac}$ as a function of $m$ in Fig.~\ref{criticalthickness}. At the $n$ interface it is almost constant in value and $\sim 0.6$~eV higher than for bulk STO. At the $p$ interface however, it decreases with increasing $m$ and becomes negative for a critical value of $m$ between 3 and 4. Thus, for $m \geq 4$ oxygen vacancies are preferentially formed at $p$ interfaces rather than in the bulk of the materials or at the $n$ interface. To better understand this behaviour, it is useful to analyze how the atomic relaxation depends on the polarity of the system.

In the absence of atomic relaxation, uncompensated internal electric fields in an LAO$|$STO multilayer (as measured for example by core level shifts) lead to the occupied LAO oxygen $p$ band rising in energy from a minimum at the $n$ interface to cross the unoccupied Ti $d$ conduction band minimum so the interface becomes conducting. When atomic relaxation is allowed, so-called {\em zigzag} buckling of the Ti-O-Ti and Al-O-Al chains in the TiO$_2$ and AlO$_2$ planes occurs \cite{Vonk:prb07,Park:prb06,Pentcheva:prl09,Schwingenschlogl:epl09,Chen:prb09,Bristowe:prb09} that compensates the electric fields and makes the whole system insulating again. However this buckling itself costs energy and as $m$ increases, oxygen vacancy formation may become a more favourable way of compensating the polarity and releasing the strain. For $p$-$n$ multilayers, vacancies at the $p$ interface will suppress the internal electric field much more strongly than at the $n$ interface, involve much less Ti-O-Ti buckling and be energetically favourable for $m$ larger than some threshold value. This argument is supported by a plot of the average relative displacements (excluding interfacial layers) of the Ti ions and O ions in the $z$ direction shown in Fig.~\ref{criticalthickness}. The buckling is clearly strongly suppressed when oxygen vacancies are formed at the $p$ interface and reaches a saturation value when the multilayer thickness is increased. However, compared to the clean interface case, it is almost unaffected by introducing oxygen vacancies at the $n$ interface.  The core level shifts are reduced from 0.9 to 0.15 V/unit cell.

{\em \color{red} Electronic structure:} A single oxygen vacancy donates two electrons to the system and the distribution and character of these excess electrons will dominate its transport properties. In a clean LAO$|$STO multilayer, the electrostatic potential at the $p$ interface is lower than at the $n$ interface so even if oxygen vacancies are generated at the $p$ interface, the excess electrons will be driven by the electrostatic potential to the $n$ interface. The charge distribution of the occupied conduction band states is plotted in Fig.~\ref{structure}. Consistent with previous studies \cite{Popovic:prl08,Salluzzo:prl09}, these have Ti $d_{xy}$-orbital character if the Ti ions are close to the interface, otherwise they consist of a mixture of Ti $d_{xz}$- and $d_{yz}$-orbital character. Though the introduction of oxygen vacancies represents a major change to the atomic structure, the change to the electronic structure is minor compared to a clean LAO$|$STO multilayer with only $n$-type interfaces.

Even though it is energetically favourable to form oxygen vacancies at a $p$ interface, the excess electrons are transferred to the $n$ interface which will tend to be conducting while the $p$ interfaces will be insulating. This asymmetry implies a spatial separation between impurity scattering and transport regions which, by analogy with proximity doping in semiconductor heterostructures, can give rise to a high mobility at LAO$|$STO interfaces. This result agrees with recent experimental data \cite{Nakagawa:natm06} which shows evidence for the presence of oxygen vacancies at the $p$ interface while the mobile carriers are at the $n$ interface. Though most experimental studies have been made \cite{Ohtomo:nat04,Nakagawa:natm06,Thiel:sc06} on samples consisting of STO substrates covered with several layers of LAO which only contain a single $n$ interface, the conclusions of this paper are qualitatively valid for the single interface case if we regard the surface of LAO as a pseudo-$p$ type interface.


{\em \color{red} Conclusion:} Using first-principles calculations, we show how oxygen vacancy formation, charge transfer and atomic relaxation in response to polar discontinuity at LAO$|$STO interfaces are strongly coupled. Oxygen vacancies are preferentially formed at $p$ rather than $n$ interfaces and the thickness-dependent formation energy provides an alternative explanation for the critical thickness observed in experiments while simultaneously explaining the failure to observe core level shifts. The conduction electrons produced when an oxygen vacancy is formed at a $p$ interface move to the $n$ interface where their interaction with the vacancies is minimal explaining the observed high mobilities.

{\em \color{red} Acknowledgments:} This work was supported by
``NanoNed'', a nanotechnology programme of the Dutch Ministry of
Economic Affairs and by EC Contract No. IST-033749 “DynaMax.”
The use of supercomputer facilities was sponsored by the
``Stichting Nationale Computer Faciliteiten'' (NCF) which is
financially supported by the ``Nederlandse Organisatie voor
Wetenschappelijk Onderzoek'' (NWO).


\end{document}